\documentclass[aip,jcp,groupedaddress, numerical bibliography]{revtex4-1}

\usepackage{graphicx}%
\usepackage{dcolumn}%
\usepackage{bm}%
\usepackage{array}%
\usepackage{amsmath}%

\begin{document}

\title{Diffusion-limited aggregation with polygon particles}

\author{Li Deng}

\author{Yanting Wang}
\email{wangyt@itp.ac.cn.}

\author{Zhong-Can Ou-Yang}

\affiliation{State Key Laboratory of Theoretical Physics, Institute of Theoretical Physics, Chinese Academy of Sciences, 55 East Zhongguancun Road, P.O.Box 2735, Beijing 100190, China}

\date{\today}

\begin{abstract}

Diffusion-limited aggregation (DLA) assumes that particles perform pure random walk at a finite temperature and aggregate when they come close enough and stick together. Although it is well known that DLA in two dimensions results in a ramified fractal structure, how the particle shape influences the formed morphology is still unclear. In this work, we perform the off-lattice two-dimensional DLA simulations with different particle shapes of triangle, quadrangle, pentagon, hexagon, and octagon, respectively, and compared with the results for circular particles. Our results indicate that different particle shapes only change the local structure, but have no effects on the global structure of the formed fractal cluster.
The local compactness decreases as the number of polygon edges increases.

\end{abstract}

%\keywords{diffusion limited aggregation, polygon particle, global and local structures}
\maketitle

\section{\label{sec1}introduction}

Various aggregation mechanisms\cite{viseck1989fractal} have been proposed to theoretically investigate disordered growth under non-equilibrium conditions, such as nanoparticle and colloidal aggregation, \cite{fukami2007general, weitz1984fractal, sorensen2001light, wang2000fractal} among which the diffusion-limited aggregation (DLA) has been intensively studied by many researchers. \cite{witten1981diffusion,jullien1987aggregation, tolman1989off,lin1990universal, meakin1999historical, menshutin2012scaling}
The original DLA model proposed by Witten and Sander \cite{witten1981diffusion} performs a random simulation at a finite temperature on a two-dimensional square lattice.
In this model a particle is initially fixed at the origin, and more particles are then released one by one and perform random walk in the space until they become close enough and stick on the central cluster.
The above DLA procedure generates a statistical self-similar structure whose scale-invariant properties can be described by the fractal geometry. \cite{mandelbrot1983fractal, meakin1983formation, kolb1983scaling}
Power law scaling of the two-point correlation function was discovered in the initial work by Witten and Sander \cite{witten1981diffusion} and the fractal dimension of a cluster formed on a two-dimensional lattice is $5/3$ regardless of the lattice geometry. \cite{viseck1989fractal}

DLA is a successful abstract model for qualitatively understanding irreversible aggregation of ramified fractal structures observed in many experiments. \cite{brechignac2003thermal,miyashita2005fractal,praud2005fractal,lando2006coarsening, tang2009fractal}
However, there still exist many other experimental observations which beyond the explanation given by DLA.
For instance, aggregation can result in regular dendrite fractal structures, such as magnetic $\alpha$-$\rm{Fe}_{2}\rm{O}_{3}$, \cite{cao2005single} fractal assembly of copper nanoparticles, \cite{ji2008replacement} and the snowflake structure in nature, \cite{bentley1962snow} which may have direct connections with the experimental observations \cite{tang2009fractal, liu2000nucleation} that particle shape plays an important role in the formation of those structures.
Consequently,  some simulations \cite{liu2000nucleation, mohraz2004effect, menshutin2010morphological} have studied the effect of particle anisotropy on DLA morphology.
Liu $et\ al.$ \cite{liu2000nucleation} studied the influence of the monomer anisotropy to the DLA structure and they concluded that anisotropic monomers still lead to fractal patterns.
Mohraz $et\ al.$ \cite{mohraz2004effect} investigated colloidal rod aggregation in three dimensions by both experiments and simulations and found that the fractal dimension increases with increasing rod aspect ratio.
Menshutin and Shchur \cite{menshutin2010morphological} found that different degrees of monomer anisotropy result in clusters with different fractal dimensions and the noise-reduction level can change the morphology of the clusters.
Nevertheless, no studies have been done to investigate the influence of particle shape to the formed morphology of DLA.

In this work, we perform the two-dimensional off-lattice DLA simulations with different particle shapes of triangle, quadrangle, pentagon, hexagon, and octagon, respectively, and analyze the local and global properties of the finally formed fractal structures, and compare with the results for circular particles.
Our results indicate that different polygon particles lead to different local structures, but they have negligible effect on the global behavior of the fractal structure. In addition,
the local compactness decreases as the number of polygon edges increases.
The paper is organized as follows: the simulation and analysis methods are described in Section \ref{sec2}, the results are shown in Section \ref{sec3}, followed by conclusions given in Section \ref{sec4}.

\section{\label{sec2}methods}

In this section, we describe our DLA simulation method, the global structure analysis methods, the skeleton algorithm identifying the main branches of fractal structure, and the orientational order parameter calculation characterizing the local structure.

\subsection{Simulation method}
Our two-dimensional DLA simulations with circular particles are the same as the original one by Witten and Sander \cite{witten1981diffusion} except that we perform off-lattice instead of on-lattice simulations.
Additional simulations were conducted with one of the five polygons as the particle shape: triangle, quadrangle, pentagon, hexagon, and octagon.
The radius of the circle is $a$ and all the polygons are regular ones whose circumscribed circle has the same radius of $a$.

In all simulations, a seed particle is initially fixed at the origin.
At each step, a free particle is released at a random position with a distance $d=d_{\rm f}+d_0$ from the seed particle, where $d_{\rm f}$ is the distance between the seed particle and the outer-most particle on the cluster and $d_0$ is the distance between the free particle and the outer-most particle on the cluster.
The released free particle is allowed to translate in any directions with a random displacement generated from a uniform distribution$[-d_{\rm t},d_{\rm t}]$ and to rotate with a random angle generated from a uniform distribution $[-\delta_{\rm r},\delta_{\rm r}]$.

%*****************
%
After each movement, the distance $d_{\rm n}$ between the free particle and the nearest particle on the cluster is calculated.
The free particle stops moving only if $d_{\rm n}$ is smaller than a critical distance $d_{\rm c}$, and in the circular case, its position is adjusted so that its center has a distance of $2a$ from the center of the nearest particle.
For polygons, the orientation is also adjusted along with the distance so that its edge overlaps with the closest edge of the nearest particle.
After the free particle sticks on the cluster, a new free particle is released and the above procedure repeats.

The distance $d_0$ between an outer-most particle on the cluster and the initial position of a released particle is related to the particle concentration of a real system.
The cutoff $d_{\rm c}$ in the simulation corresponds to the effective range of the adhesive interaction between particles.
The amplitudes of $d_{\rm t}$ and $\delta_{\rm r}$ reflect the system temperature.
The periodic boundary condition \cite{frenkel2002understanding} was applied to avoid the escape of particles from the simulation space, and the neighbor list algorithm \cite{frenkel2002understanding} was adopted to accelerate the simulations.
In this work, we set $a=1$, $d_{\rm c}=3$, $d_{\rm t}=0.5$, $\delta_{\rm r}=0.5$ in radian, the simulation box to be a square with the side length $L=1000$, and $d_0=100$ to make sure that the DLA process is in a low concentration condition.
Nine independent runs have been performed and each run contains $M=10000$ particles.

\subsection{Fractal dimension}
Historically, the fractal dimension of DLA was initially evaluated by Witten and Sander\cite{witten1981diffusion} through the radial distribution function (RDF) $g(r)$ of the cluster.
According to the statistical self-similar property of a DLA cluster, the RDF has the form $g(r)\propto{r^{D_{\rm f}-D}}$ where $D_{\rm f}$ is the fractal dimension of the cluster and $D$ is the dimension of the Euclidean space in which the cluster is embedded.
Later on, most experiments and simulations have utilized a more convenient relation between the number of particles and the distance from the origin to determine the fractal dimension $D_{\rm f}$. \cite{menshutin2006test}
The number $N(l)$ of particles inside a circle with radius $l$ and centered at the origin can be written as\cite{heinson2010does, oh1998structure}
\begin{equation}
\label{df1}
N(l)={k_{0}\times{(l/a)^{D_{\rm f}}}},
\end{equation}
where the prefactor $k_0$, related to the lacunarity of the cluster,\cite{malcai1997scaling} is different for clusters formed by different aggregation mechanisms. \cite{oh1998structure}
Oh and Sorensen \cite{oh1998structure} argued that the relation is valid only when $l$ is larger than a critical length $l_{\rm c} \sim 10a$.
According to Eq.~\ref{df1}, a fractal cluster with a larger $D_{\rm f}$ has more particles within the same radius from the origin, so the fractal dimension $D_{\rm f}$ can be used to characterize the global compactness of a cluster.
The prefactor $k_0$ can be used to characterize the local compactness of a cluster, as illustrated by our later local structure analysis.

\subsection{Skeleton algorithm}
The skeleton of a cluster can be computationally identified to study its large scale properties, \cite{schwarzer1996number} and from the skeleton of a cluster we can easily obtain the number of main branches $n_{\rm b}$.
The main branches are those branches whose length, which is the number of particles from the tip of the branch to the origin, is larger than a critical length $L_{\rm c}$, which is comparable to the cluster size.

\begin{figure}
\includegraphics{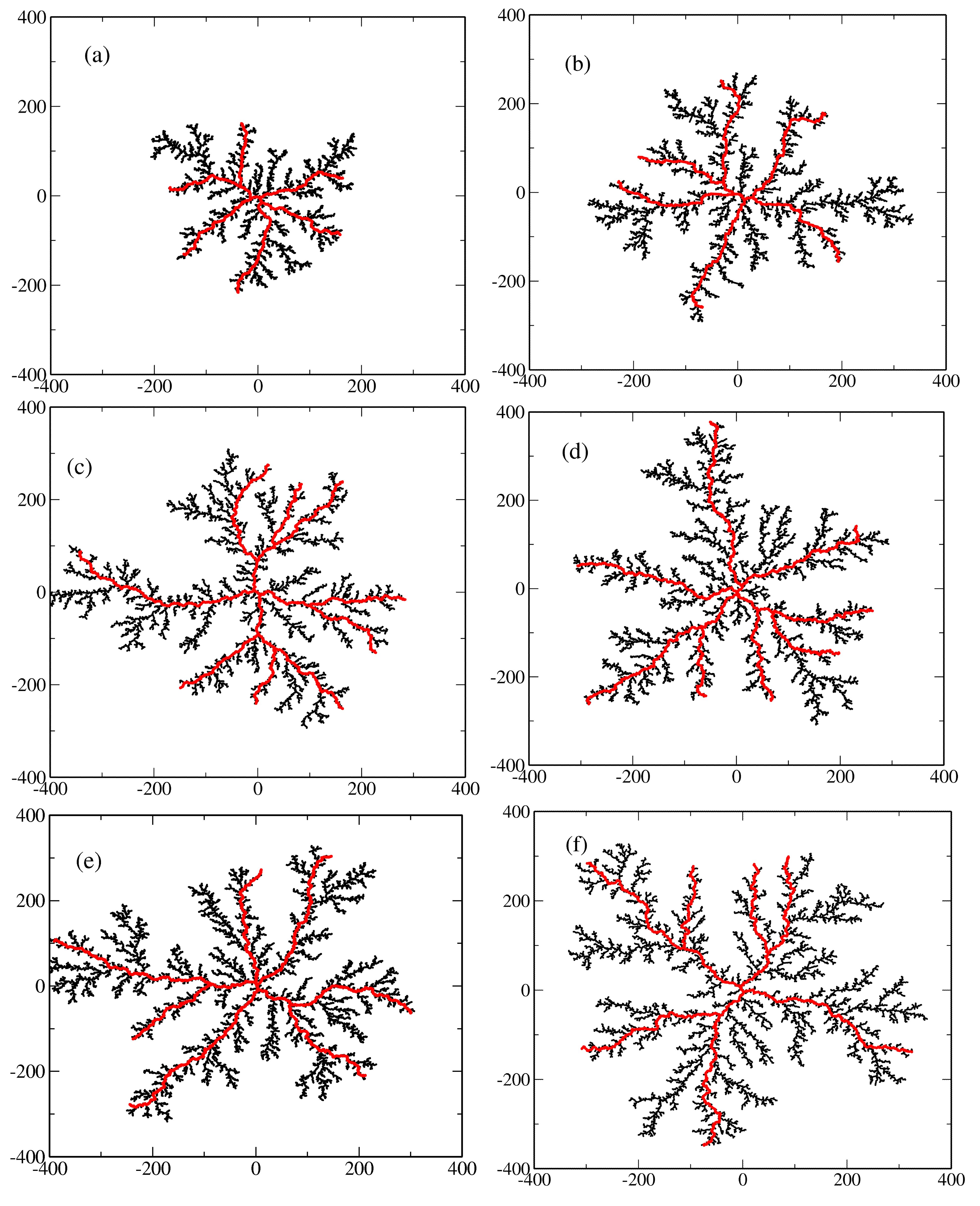}
\caption{\label{skeleton} DLA clusters with different polygon particles of $a)$ triangle, $b)$ quadrangle, $c)$ pentagon, $d)$ hexagon, $e)$ octagon, and $f)$ circle. The red lines depict the calculated skeletons.
}
\end{figure}

In a cluster, there exist many branches whose length is comparable to the cluster size, but not all those branches are intrinsically different because some of them share lots of particles and only differ in a few particles.
If the number of shared particles exceeds an critical value $S_{\rm c}$, all these branches are regarded as in the same main-branch class and only one of them is chosen to represent this class.
According to the algorithm developed by Schwarzer $et\ al.$ \cite{schwarzer1996number} for the DLA cluster growth process, there exists a parent-child relation between two neighbor particles, in which the particle joining the cluster later is called "child" and the particle joining the cluster earlier is called "parent".
By applying this relation repeatedly we can identify all the branches from a tip particle to the seed particle at the origin.
After all branches of the cluster are identified, we can obtain the main branches of the cluster as follows:
1) the branches whose length is larger than a critical branch length $L_{\rm c}$ are picked up from all the tip branches;
2) the similarity $S$, defined as the number of particles shared by two branches, are calculated for each pair of branches;
3) the branches with $S$ larger than a critical similarity $S_{\rm c}$ can be considered as belonging to the same main-branch class and only one of them is chosen randomly to represent this class.
The skeletons obtained with the above algorithm for the DLA clusters with different polygon particles are drawn in Fig.~\ref{skeleton}. We can see that the selected main branches are not necessarily the longest in a main-branch class since they were randomly chosen.

\graphicspath{{./picture/}}

With the main branches of the cluster determined, we then calculate the angle $\theta$ between two neighboring main branches and the direction of each main branch is determined by fitting the main branch with a straight line.
The distribution of angle $\theta$ quantifies the rotational symmetry of the skeleton.
The direction of the fitted straight line is mainly determined by the particles shared by all the branches in the same class, so the random selection of the main branches does not influence noticeably the calculated rotational symmetry.

\subsection{Orientational order parameter}
In contrast to the fractal dimension defined to quantify the global structure, the orientational order parameter is used to characterize the local symmetry of a DLA cluster.
The $n \rm{th}$ orientational order parameter $\Phi_n$ is defined as \cite{gotcheva2005continuous}
\begin{equation}
\label{or1}
\Phi_{n} = \frac{1}{N_{\rm t}}\sum_{i=1}^{N_{\rm t}}{\frac{1}{z_i}\sum_{j=1}^{z_i}{e^{in\theta_{ij}}}},
\end{equation}
where $N_{\rm t}$ is the total number of particles in the cluster, $z_i$ is the number of nearest neighbors of particle $i$, and $\theta_{ij}$ is the angle of the vector $\vec{r}_{ij}$ 
from particle $i$ to particle $j$ with respect to a fixed vector $\vec{a_0}$, which was chosen in our calculation to be the unit vector parallel to the $y$ axis.
According to the definition, $\Phi_{n}$ equals to $1$ if the local structure has a perfect $n$-fold symmetry.
To observe the change of the orientational symmetry with respect to the length scale, we extend the definition of the orientational order parameter to be distance-dependent:
\begin{equation}
\label{or2}
\Phi_{n}(r) = \frac{1}{N_{\rm t}}\sum_{i=1}^{N_{\rm t}}{\frac{1}{z_i^{'}}\sum_{j=1}^{z_i^{'}}{e^{in\theta_{ij}}}}.
\end{equation}
Eq.~\ref{or2} only differs from Eq.~\ref{or1} in the number of particles in the second sum $z_i^{'}$, which is now over the particles in the range $r_{ij} < r$, rather than all the nearest particles.
With the extended orientational order parameter, we can analyze the cluster geometry at different length scales.

\section{\label{sec3}results and discussion}

\subsection{Fractal dimension}
\begin{figure}
\includegraphics{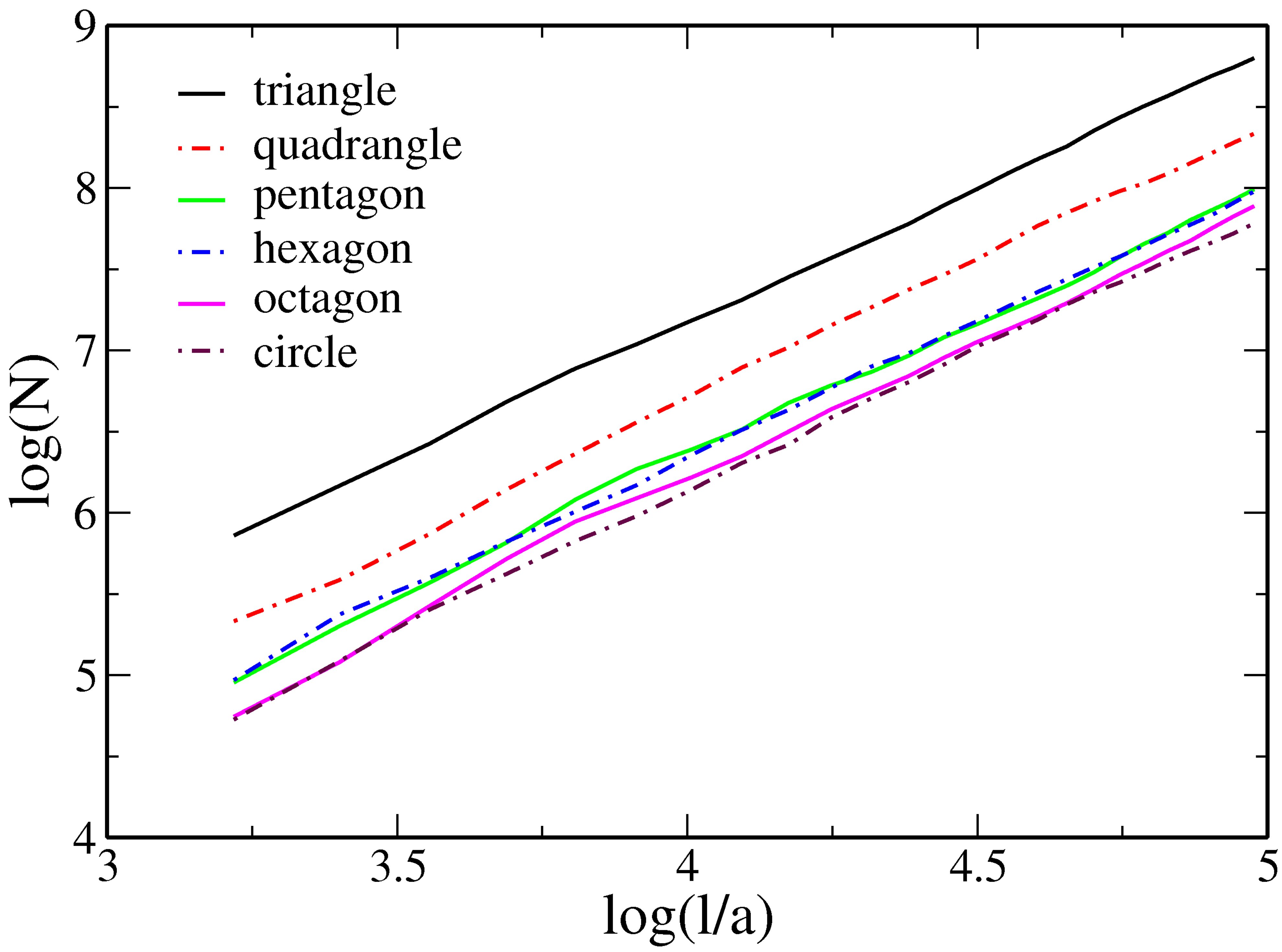}
\caption{\label{fracD} Log-log plot of the number of particles $N$ as a function of the normalized distance $l/a$ from the origin.}
\end{figure}

In this section, we present our calculation of the fractal dimensions for the DLA clusters formed by different polygon particles through the relation between $N(l)$ and $l$ defined in Eq. \ref{df1}.
The particle numbers $N(l)$ versus the normalized radius from the origin $l/a$ are shown in a log-log plot in Fig.~\ref{fracD} with only the linear part ranging from $25a$ to $148a$.
All the lines in Fig.~\ref{fracD} are parallel to each other and the interceptions of these lines decrease with increasing number of particle edges. According to Eq.~\ref{df1}, the slopes of these lines correspond to the fractal dimensions of the clusters and the interceptions correspond to the prefactors $k_0$. The fitted fractal dimensions and their standard deviations are listed in Tab.~\ref{table1}. We can see from both Fig.~\ref{fracD} and Tab.~\ref{table1} that the clusters formed by different polygon particles have similar fractal dimensions but different prefactors.

\graphicspath{{./picture/}}

\begin{table}[h]
\caption{\label{table1} Fractal dimensions $D_{\rm f}$ and their standard deviations $\sigma$.}
\begin{tabular}{l c r}
Particle  & $D_{\rm f}$ & $\sigma$\\
\hline
triangle & 1.67 & 0.04\\
quadrangle & 1.69 & 0.07\\
pentagon & 1.69 & 0.07\\
hexagon & 1.69 & 0.04\\
octagon & 1.75 & 0.08\\
circle & 1.73 & 0.07\\
\end{tabular}
\end{table}

In Tab.~\ref{table1}, the DLA clusters with triangle, quadrangle, and hexagon particles have the same fractal dimension of about $1.69$, consistent with the fractal dimension of $5/3$ for the on-lattice simulations performed on a triangle, square, or honeycomb lattice.\cite{viseck1989fractal}
Furthermore, our results show that other particle shapes of pentagon, octagon, and circle, which have no corresponding lattice structures, also have similar fractal dimensions.
This result indicates that the shape of particles has no noticeable influences on the global structure of a two-dimensional DLA cluster, but influences significantly the prefactor $k_0$ defined in Eq.~\ref{df1}, which characterizes the local compactness of the cluster.
The prefactor $k_0$ will be studied in detail in subsection \ref{sec3:D}.

\subsection{Cluster skeleton}
As described by Schwarzer $et\ al.$,\cite{schwarzer1996number} the number of main branches $n_{\rm b}$ obtained by the skeleton algorithm is also an important property characterizing the global structure of the cluster.
They have also shown that, for the DLA cluster with circular particles in two dimensions, the number of main branches takes a constant value of $n_{\rm b}=7.5\pm 1.5$, independent of the cluster size.
In this work, we obtained the average values of $n_{\rm b}$ for triangle, quadrangle, pentagon, hexagon, octagon, and circle, respectively, as listed in Tab.~\ref{table2}, which are all close to the value $7.5 \pm 1.5$ reported in Ref. 31.

\begin{table}[h]
\caption{\label{table2} Number of main branches $n_{\rm b}$ and their standard deviations $\sigma$. }
\begin{tabular}{l c r}
Particle  & $n_{\rm b}$ & $\sigma$\\
\hline
triangle & 6.89 & 1.05\\
quadrangle & 6.22 & 1.09\\
pentagon & 6.89 & 1.05\\
hexagon & 6.75 & 1.28\\
octagon & 7.25 & 0.71\\
circle & 6.78 & 0.97\\
\end{tabular}
\end{table}

\begin{figure}
\includegraphics{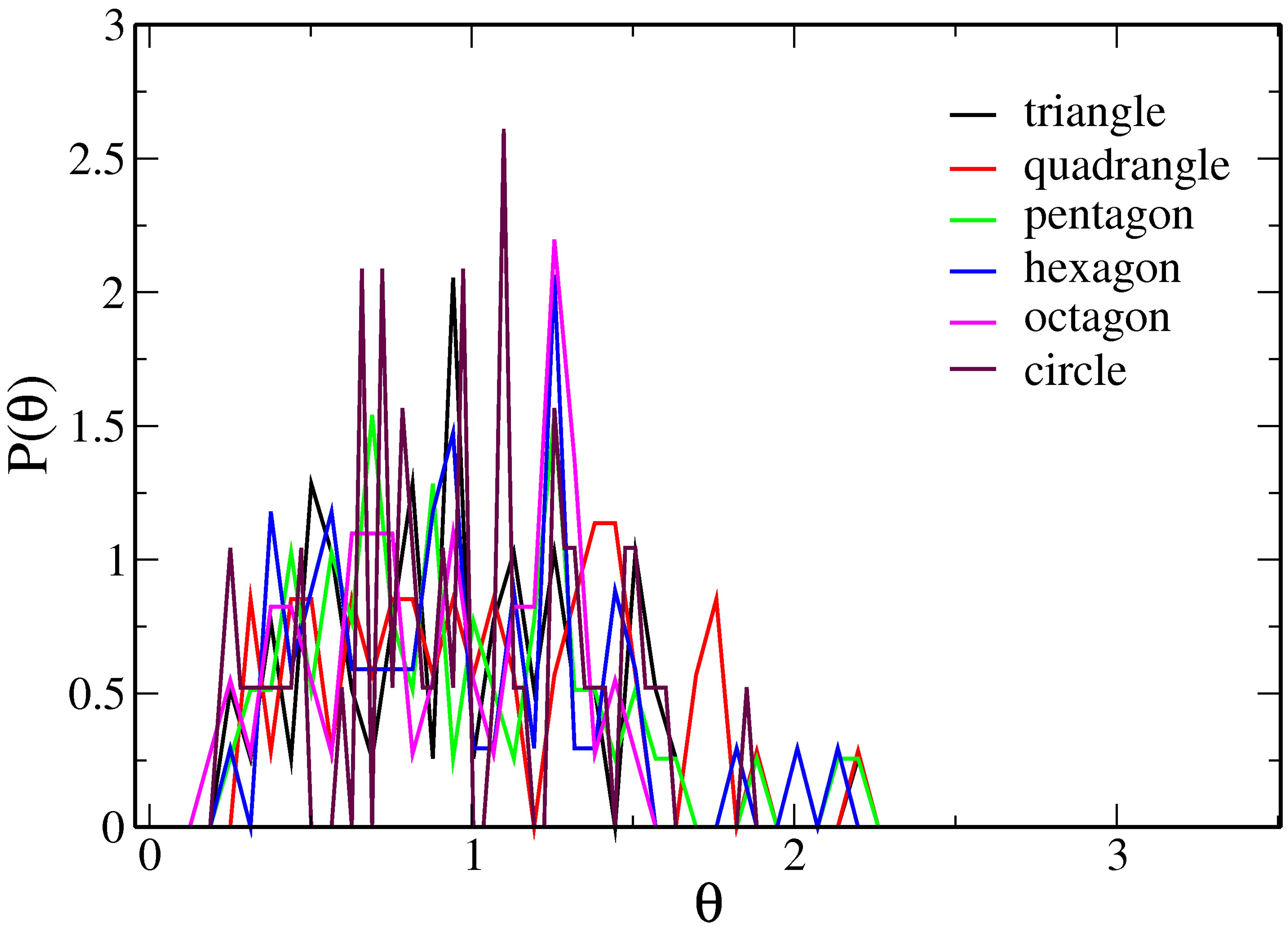}
\caption{\label{angle-dis} Distributions of angle $\theta$ between two neighboring main branches in the skeletons of the DLA clusters with various particle shapes. }
\end{figure}

Schwarzer $et\ al.$ \cite{schwarzer1996number} has shown that, in a two-dimensional DLA, the increasing rate of free space is the same as the increasing rate of space screened by the dangling branches aside the main branches, so the number of main branches is a constant during two-dimensional aggregation.
Consistently, our calculated skeletons shown in Fig.~\ref{skeleton} have similar structures for different polygon particles.
The distribution of angle $\theta$ between two neighboring main branches was then calculated to characterize the rotational symmetry of the skeleton. Fig.~\ref{angle-dis} indicates that the distributions of angle $\theta$ are similar for all the clusters formed by various polygon particles, and the average angle values are listed in Tab.~\ref{table3}. All these results indicate that the particle shape has no effect on the skeleton of DLA cluster, neither on the number nor on the structure of the main branches.

\graphicspath{{./picture/}}

\begin{table}[h]
\caption{\label{table3} Angle $\theta$ between two neighboring main branches and their standard deviations $\sigma$.}
\begin{tabular}{l c r}
Particle  & $\theta$& $\sigma$\\
\hline
triangle & 0.91 & 0.40\\
quadrangle & 1.01 & 0.46\\
pentagon & 0.91 & 0.44\\
hexagon & 0.93 & 0.43\\
octagon & 0.87 & 0.37\\
circle & 0.93 & 0.40\\
\end{tabular}
\end{table}

\subsection{Orientational order parameter}
\begin{figure}[h!]
\includegraphics{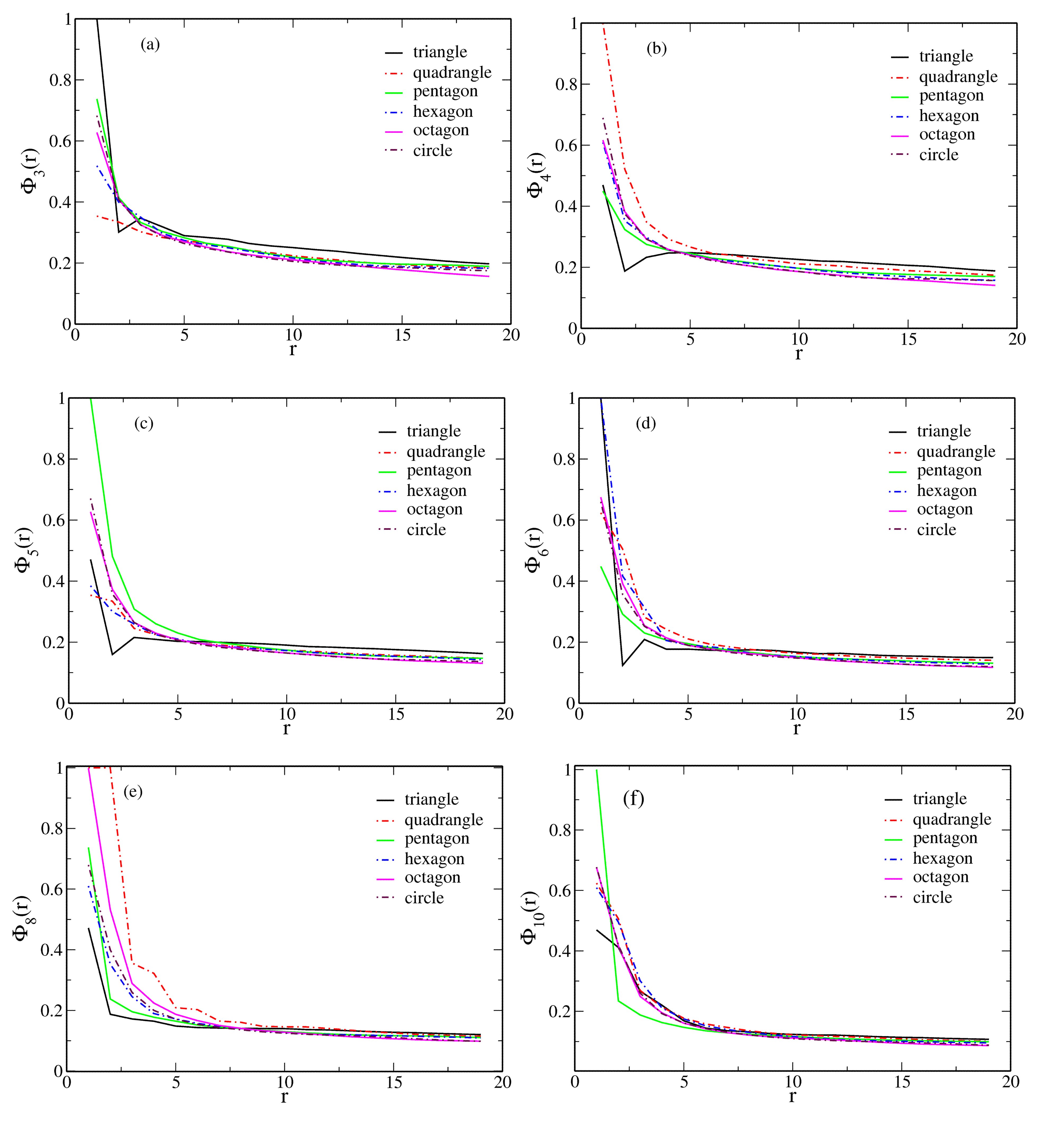}
\caption{\label{orientation} Orientational order parameters with different orders versus radius $r$ for the clusters formed by particles with various shapes.(a)Three-fold, (b)four-fold, (c)five-fold, (d)six-fold, (e)eight-fold, and (f)ten-fold. }
\end{figure}

The local structure of the DLA cluster is important for studying the effect of different physical conditions on the growth process.
Mandelbrot \cite{mandelbrot1992plane} studied in detail the lacunarity distribution of cluster, which is related to the compactness of the cluster, and found that the lacunarity distribution is different at different length scales.
In our work, the orientational order parameter was calculated to quantify the local structure of the DLA cluster.
We calculated with Eq.~\ref{or2} six orientational order parameters: \cite{gotcheva2005continuous} $\Phi_3, \Phi_4, \Phi_5, \Phi_6 , \Phi_8$, and $\Phi_{10}$ as a function of distance for all simulations.
The order parameters $\Phi_n$ as a function of distance $r$ are shown in Fig.~\ref{orientation}.

\graphicspath{{./picture/}}

As can be seen from Fig.~\ref{orientation}, the value at the smallest $r$ of each curve for the $n$th orientational order parameter is always $1$ for a polygon with the $n$-fold symmetry because the nearest neighbors of a particle are always arranged with the orientation determined by the polygon shape.
All figures in Fig.~\ref{orientation} indicate that, at the length scale $a<l<5a$, the orientational symmetry is apparently different for the clusters formed by different polygons, but at the length scale $l>10a$, all clusters have the same isotropic orientational symmetry.
The local structure of a cluster is determined by the competition between the randomness due to diffusion and the orientational symmetry due to polygon shape around a particle.
The influence of polygon shape on a cluster's structure decreases rapidly with increasing distance.
As shown in Fig.~\ref{orientation}, when $l<5a$, the particle alignment is mainly determined by the particle geometry; when a DLA cluster is growing, the number of new growth sites which can accept new particles increases exponentially and distribute isotropically around the particle at the origin, so when the length scale $l>10a$, the randomness due to diffusion dominates and the cluster structure is independent of the particle geometry.

\subsection{\label{sec3:D}Local compactness and prefactor}

The above results for orientational order parameters indicate that the particle shape only affects the local structure of the cluster, and in Fig.~\ref{fracD} it is shown that the prefactors $k_0$ are different for different particle shapes.
Therefore, there should exists a direct relation between the local structure and the prefactor $k_0$.
According to the original definition of fractal structure, \cite{viseck1989fractal} the fractal dimension $D_{\rm f}$ is calculated by measuring the volume of fractal structures embedded in a $D$-dimensional Euclidean space.
The volume of fractal structures is obtained by counting the number $M(r)$ of balls with radius $r$ needed to cover the fractal structure. For an ideal fractal structure, the self-similarity is satisfied in all the length scale, \cite{viseck1989fractal} so the numbers of counting balls with radius $r_1$ and $r_2$, respectively, should satisfy the following relation
\begin{equation}
\label{recursion1}
M(r_2)=M(r_1)\times{(r_1/r_2)^{D_{\rm f}}},\       r_1>r_2.
\end{equation}
Form Eq.~\ref{recursion1} we can define the average number of balls with radius $r_2$ in a ball with radius $r_1$ as
\begin{equation}
\label{recursion2}
M(r_2;r_1)=\frac{M(r_2)}{M(r_1)} ={(r_1/r_2)^{D_{\rm f}}},\       r_1>r_2.
\end{equation}
The above relation can be easily extended recursively to the case with three radii
\begin{equation}
\label{recursion3}
M(r_3;r_1)=M(r_2;r_1)\times M(r_3;r_2),\      r_1>r_2>r_3.
\end{equation}
The quantity $M(a;l)$ is equivalent to $N(l)$ defined in Eq.~\ref{df1}. A detailed study by Oh and Sorensen\cite{oh1998structure} concludes that the self-similar relation does not satisfy at all length scales,
rather there exists a critical length scale of about $10a$, when $l>10a$, the DLA cluster has different self-similar property for different aggregation kinetics, but no differences were found for $l<10a$.
Recently, Heinson $et\ al.$ \cite{heinson2010does} studied the inertia tensor of many different fractal clusters and pointed out that the prefactor $k_0$ is related to the cluster's morphology. In addition,
our results for the orientational order parameter indicate that the particle shape only affects the cluster structure at finite length scales.
These results suggest that there should exist a cutoff distance $l_{\rm c}$ for the recursive relation in Eq.~\ref{recursion3}, and the relation between the number of particles and the distance from the origin can be written as
\begin{equation}
\label{recursion4}
N(l)=N(l_{\rm c})\times M(l_{\rm c};l) = N(l_{\rm c})\times{(l/l_{\rm c})^{D_{\rm f}}},
\end{equation}
where $N(l)$ and $N(l_{\rm c})$ are the numbers of particles within radii $l$ and $l_{\rm c}$, respectively, as defined in Eq.~\ref{df1}. Compared with Eq.~\ref{df1}, we obtain the relation between $k_0$ and $N(l_{\rm c})$ as
\begin{equation}
\label{nl-k}
k_0=N(l_{\rm c})\times a^{D_{\rm f}}/l_{\rm c}^{D_{\rm f}}.
\end{equation}
By fitting the simulation results, we found that Eq.~\ref{nl-k} is roughly satisfied for all cases when $l_{\rm c}=5a$. The data of $\ln(N(l_{\rm c}=5a) \times a^{D_{\rm f}}/l_{\rm c}^{D_{\rm f}})$ and $\ln(k_0)$ are compared in Tab.~\ref{table4}. The prefactor $k_0$ for each case was determined by the linear fitting of $N(l)$ $vs$ $l$ averaged over nine independent simulations. It is clear that $k_0$ corresponds to the local compactness of the cluster when $l<5a$ and the local compactness decreases as the number of polygon edges increases.
These results are in agreement with Heinson $et\ al.$'s conclusion \cite{heinson2010does} that $k_0$ characterizes the anisotropy of the cluster which only exhibits in local structures.

\begin{table}[h]
\caption{\label{table4} Prefactor values for the cluster with various particle shapes.}
\begin{tabular}{l c r}
Particle  & $\ln(N(l_{\rm c}=5a)\times a^{D_{\rm f}}/l_{\rm c}^{D_{\rm f}})$& $\ln(k_0)$\\
\hline
triangle & 0.312 & 0.464\\
quadrangle & -0.233 & -0.114\\
pentagon & -0.326 & -0.335\\
hexagon & -0.465 & -0.398\\
octagon & -0.738 & -0.769\\
circle & -0.79 & -0.82\\
\end{tabular}
\end{table}

\section{\label{sec4}conclusion}
In conclusion, we have done a series of two-dimensional off-lattice DLA simulations with particle shapes of different polygons, and the global and local structures of the clusters have been studied in detail.
By analyzing the fractal dimension, we conclude that the geometry of the particles has no effects on the global structure of cluster.
We have also shown that the DLA clusters formed by different polygon particles have the same number of main branches and the same skeleton symmetry.
The results of the extended orientational order parameters show that the particle shape influences the local structure of the clusters at a finite length scale, but the effects decay quickly at larger length scales.
The prefactor connecting the number of particles and the radius from the origin quantifies the local compactness of the DLA clusters, and our results for the prefactor indicate that the local compactness decreases as the number of polygon edges increases. Therefore, we conclude that the particle shape only affects the local structure of a two-dimensional DLA cluster, but has no effects on its global structure.

\begin{acknowledgements}
This research was supported by the Hundred Talent Program of the Chinese Academy of Sciences (CAS) and the National Natural Science Foundation of China Nos. 10974208, 11121403, 1083401401, and 91027045. The authors thank the Supercomputing Center in the Computer Network Information Center at the CAS for allocations of computer time.

\end{acknowledgements}

\bibliography{cite}

\end{document}